# Queered Science & Technology Center: Volume 1

Sergio Carbajo*[1-5]

[1]Electrical & Computer Engineering Department, University of California, Los Angeles
[2]Physics & Astronomy Department, University of California, Los Angeles
[3]Photon Science Directorate, SLAC National Accelerator Laboratory, Stanford University
[4]California NanoSystems Institute, University of California, Los Angeles
[5]Center for Quantum Science & Engineering, University of California, Los Angeles

*Email: scarbajo@ucla.edu

**Abstract:** through the Fiat Lux seminar series at the University of California, Los Angeles, Volume 1 presents the first pedagogical and academic plans to destabilize hegemonic logics and hierarchies in STEM studies and knowledge production.

INTRODUCTION

The Queered Science & Technology Center (QSTC) at the University of California, Los Angeles was established in 2022 by Prof. Carbajo to focus on new frameworks that address overarching issues of diversity and critical representation in STEM (science, technology, engineering, and mathematics) through queer, radical feminist, and black analyses of the impact of science & technology in society. The QSTC employs these critical frameworks to destabilize sexual, gendered, racialized, anthropocentric, and able-bodies logics and hierarchies in challenging and rethinking knowledge production, as a scientific exercise and introduces new methodological resources for critical interdisciplinarity in traditional STEM studies.

The QSTC is partnered with non-profit academic, public, and private institutions, and works collaboratively with county and state, and federal-level sponsored programs tailored to promote equity in STEM fields through action in distinct areas of sciences and engineering. The Center also supports the recruitment of outstanding faculty, staff, and students; a diverse intellectual and educational community; research in STEM education; fostering new interdisciplinary connections across campus; and the empowerment of (future) STEM workforce, particularly those from underrepresented backgrounds, to affect social change that is representative of the public's interests.

METHODS: Volume 1

The QSTC will launch its first instructional and research activity through a Fiat Lux seminar series entitled "Upending the Hard Sciences." As a cornerstone of UCLA's innovative undergraduate curriculum, the Fiat Lux program offers up to 200 seminars annually. These seminars provide students and Senate faculty with small group settings to engage in meaningful discussions on a range of topics. Students receive one unit of academic credit (Pass/No Pass) and Senate faculty members from across campus can share with undergraduates their areas of intellectual passion and expertise.

*Upending the Hard Sciences* is designed to support the co-development of a sustainable, critically representative STEM-community engagement with the arts, humanities, and social sciences. The series course will consist of six consecutive panel discussions followed by Q&A, each of which will explore intersecting frameworks of knowledge and technology production and its linkage to broad social and local community (in)equities. This seminar series will examine how to destabilize hegemonic logics and hierarchies in STEM under some of the following frameworks: queer and black geographies[1,2], gender essentialism[3,4], colonial sexuality[5,6], queer ecology[7] and eco-feminism[8], non-essentiality[9], (Latinx, black) futurity[10,11], indigenous temporalities[12] and queer time[13].

## IMPACT

The STEM workforce represents approximately 23% of the total U.S. workforce today[14]. Unemployment is consistently lower among the STEM workforce compared to the non-STEM labor force, a pattern that persists even during the COVID-19 pandemic. STEM workers also have substantially higher median earnings ($55,000) than non-STEM workers ($33,000) in the US. Foreign-born workers account for 45% STEM workforce with doctoral degrees, the highest skilled STEM labor force in mathematics, computer science, physics, life sciences, social sciences, and engineering. Women and racial minorities (URM) are consistently underrepresented accounting for 34% and 30% of active STEM workers[15].

Beyond representation, academic research and scientific enterprises today are shaped by constructs of property, sovereignty, and personhood. Their relation undergirds Western societal structure, morality, and economic and justice systems. The construct of personhood in modern Western society is born out of relation to property and labor under a sovereign authority. The societal structures of expansive colonialism echo a jurisprudential framework based on a structure of domination and endless resource extraction. This is the legal basis under which exploitation and violence in neoliberalism are woven into the Western social structure and by association, into STEM research and pedagogy through Academic Industrial Complex.

Critically challenging academic and scientific frameworks that inform the U.S. (future) STEM workforce and its value systems can have a profound global impact. In total, STEM supports 69% of the U.S. GDP and $2.3 trillion in annual federal tax revenue[16]. This seminar series will interrogate the interwoven architecture of colonialism in academic STEM research and destabilize the inertia carried by market prioritization logics to formulate Western knowledge and construct models of so-called technological advantage and prosperity.